\documentclass[a4paper,11pt]{article}
\usepackage{pos}
\usepackage{lineno}

\usepackage{xspace} 
\usepackage{upgreek}

\def\lhcb   {\mbox{LHCb}\xspace}

\def\besiii {\mbox{BESIII}\xspace}
\def\cleo   {\mbox{CLEO}\xspace}
\def\CP                {{\ensuremath{C\!P}}\xspace}

\newcommand{\tev}{\aunit{Te\kern -0.1em V}\xspace}

\def\PK      {\ensuremath{K}\xspace}
\def\kaon    {{\ensuremath{\PK}}\xspace}
\def\KS      {{\ensuremath{\kaon^0_{\mathrm{S}}}}\xspace}

\title{CKM $\gamma$ measurements at LHCb}

\author*[a]{Lei Hao}

\onbehalf{on behalf of the \lhcb collaboration}
\affiliation[a]{University of Chinese Academy of Sciences,\\
  Yuquan Rd, Beijing, China}

\emailAdd{hao.lei@cern.ch}

\abstract{The tree-level measurement of the CKM $\gamma$ angle is a crucial test of $\CP$ violation in the Standard Model (SM). Discrepancies between direct measurements (tree-level decays) and indirect measurements (loop decays) could indicate physics beyond the SM. Recent measurements using decays such as $B^0 \rightarrow DK^{*0}, B^{\pm} \rightarrow [D\pi^0/\gamma]_{D^*}h^{\pm}$ (where $D$ decays to $\KS h^+h^-$ with $h=K, \pi$) and $B^{\pm} \rightarrow [h'^+h'^-\pi^+\pi^-]_Dh^{\pm}$ are presented. Additionally, the combination of previous $\gamma$ measurements at \lhcb, excluding the aforementioned results, is also presented. The \lhcb result, with a precision of $(63.8^{+3.5}_{-3.7})^{\circ}$ establishes itself as a dominant measurement in this field.}

\FullConference{The European Physical Society Conference on High Energy Physics (EPS-HEP2023)\\
 21-25 August 2023\\
Hamburg, Germany\\}

\begin{document}
\maketitle

\section{Introduction} 
The CKM matrix elements, denoted as $V_{ij}$, represent the strength of flavour-changing weak interactions~\cite{PhysRevLett.10.531}. 
\CP-violation arises from the complex phases of these CKM elements~\cite{Rosner:1996yb,ParticleDataGroup:2018ovx}. The CKM angle $\gamma\equiv$arg$(\frac{V_{ud}V_{ub}^*}{V_{cd}V_{cb}^*})$ is one of the least known CKM parameters.
\begin{figure}
    \centering
    \includegraphics[width=0.33\textwidth]{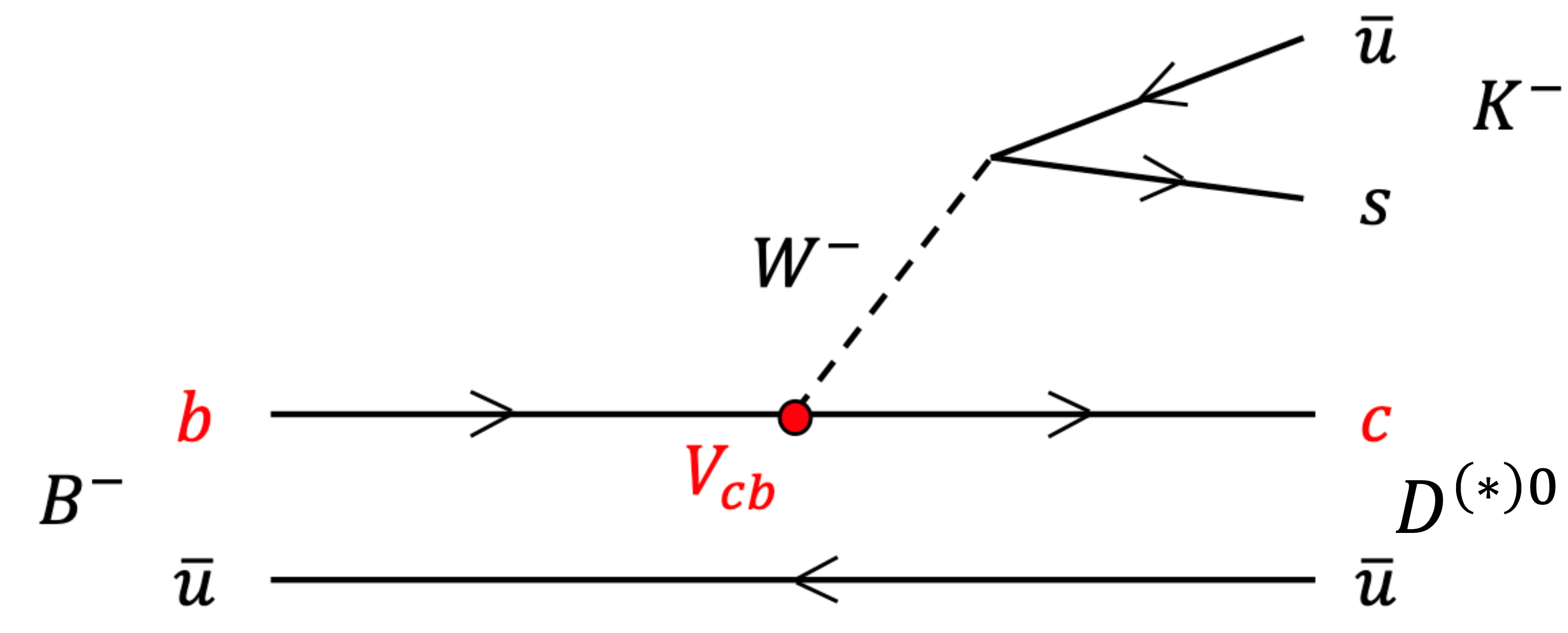}
    \includegraphics[width=0.33\textwidth]{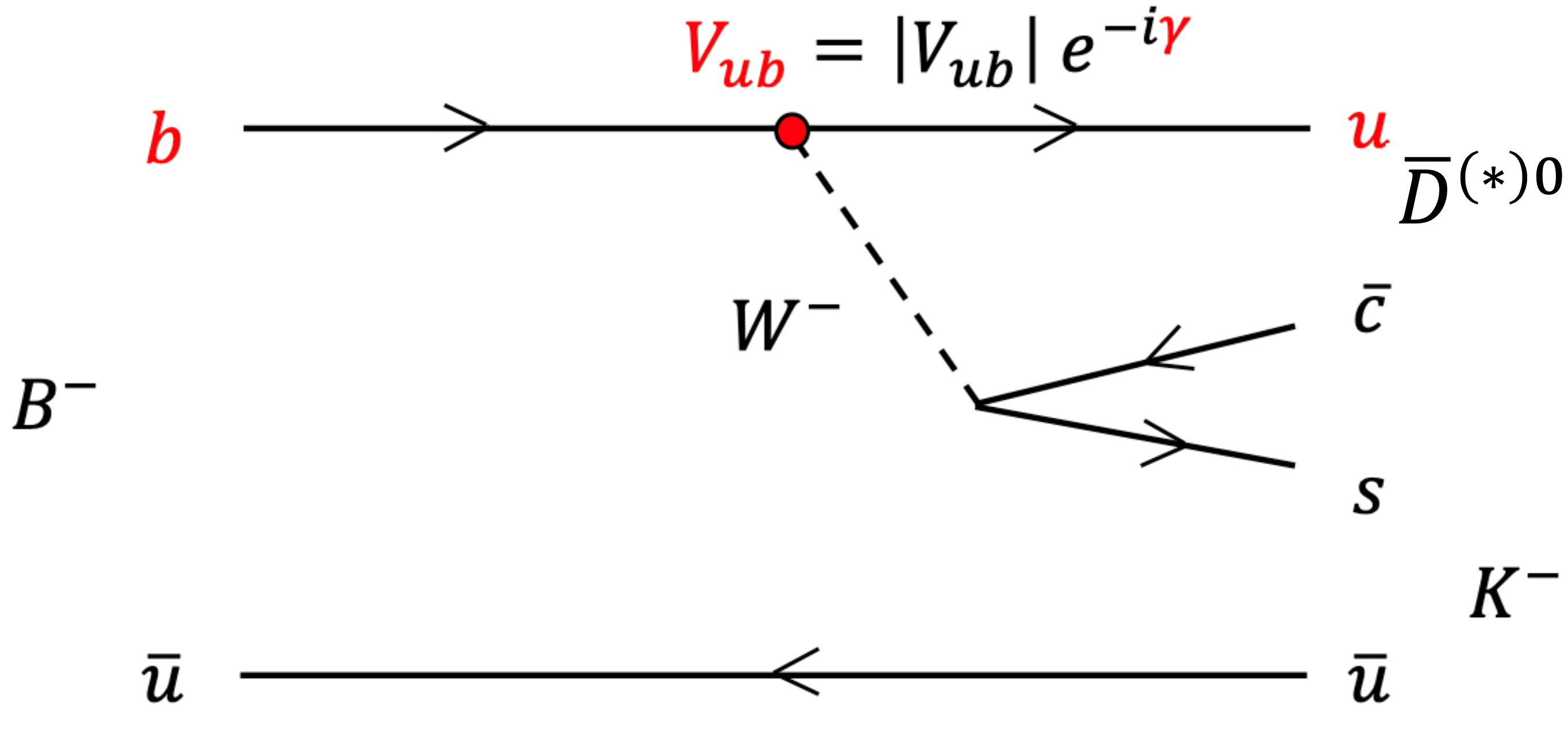}
    \caption{
    Description of the Feynman diagrams for the processes (left) $B^-\rightarrow D^{(*)0}K^-$ and (right) $B^-\rightarrow \bar{D}^{(*)0}K^-$.}
    \label{fig:triangle}
\end{figure}

Direct measurements of $\gamma$ can be accessible at tree level, so they are benchmarks of the standard model. Assuming no new physics at tree level, the theoretical uncertainties are negligible~\cite{Brod:2013sga}. Indirect measurements involve loop processes and rely on various inputs, including global fits to the unitary triangle, assuming a closed triangle. New physics can contribute to loop processes. Therefore, a discrepancy between direct and indirect measurements would be a clear indication of new physics.
 
The most powerful method for determining $\gamma$ relies on $B^{\pm}\rightarrow D^{(*)}K^{\pm}$ decays, where $D^{(*)}$ represents an admixture of the $D^{(*)0}$ and $\bar{D}^{(*)0}$ states. Fig~\ref{fig:triangle} shows the interference between $b\rightarrow c$ and $b\rightarrow u$ can give sensitivity to $\gamma$~\cite{Gronau:1991dp}. Several techniques are available for measuring $\gamma$ using decays such as $B^{\pm} \rightarrow D^{(*)}K^{\pm}$. The GLW method involves studying decays of the $D$ meson to $\CP$ eigenstates. The ADS approach requires consideration of favoured and doubly Cabibbo-suppressed $D$ decays. 
BPGGSZ utilises $D$ decays to self-conjugate final states, such as $\KS h^{+}h^{-}$, and measures the $\CP$ asymmetries over the phase space.

\section{Combination of \texorpdfstring{$\gamma$}{gamma} measurements}
The combination of $\gamma$ has been updated, including new and updated measurements with \mbox{$B^{\pm}\rightarrow [K^{\mp}\pi^{\pm}\pi^{\pm}\pi^{\mp}]_Dh^{\pm}$}~\cite{LHCb:2022nng} and \mbox{$B^{\pm}\rightarrow [h^{\pm}h'^{\mp}\pi^{0}]_Dh^{\pm}$}~\cite{LHCb:2021mmv} decays published by the \lhcb collaboration during 2022, as well as updates in charm sector~\cite{LHCb:2022awq}. The updated determination of $\gamma$ is $(63.8^{+3.5}_{-3.7})^{\circ}$, which is compatible with the previous combination by the \lhcb collaboration. Furthermore, this result is in excellent agreement with the predictions from global CKM fits~\cite{Charles:2020dfl}. 
It is the most precise determination from a single measurement to data~\cite{LHCb:2021dcr}.

\section{BP-GGSZ method}
The Dalitz plot (DP) of multi-body $D$ meson decays is divided into bins~\cite{Giri:2003ty} to measure $\gamma$ model-independently. The $D\rightarrow \KS \pi^+ \pi^-$ and $D\rightarrow \KS K^+ K^-$ modes are included. For $\KS \pi^+ \pi^-$ mode, The DP is divided into 16 bins, while for $\KS K^+ K^-$ mode, it is divided into 4 bins~\cite{LHCb:2020yot}.

The expected yields in the DP bins are related to the $\CP$ observables through Eq.~\ref{eq:yields}. The $F_i$ denotes the fractional yields of $D^0$ decays in the respective bins, which can be mainly determined from $D^{(*)}\pi$ modes. The $c_i$ and $s_i$ correspond to the cosine and sine of the strong phase difference between $D^0$ and $\bar{D}^0$ decays, respectively. They are measured from the \besiii and \cleo collaborations~\cite{BESIII:2020khq, BESIII:2020hpo, CLEO:2010iul}. Additionally, the $D^{(*)}\pi$ modes contribute to the $\gamma$ measurement and can be re-parameterised in Eq.~\ref{eq:cpodstpi} and ~\ref{eq:cpodktodpi}~\cite{GarraTico:2018nng}.
\begin{equation}
    \begin{split}
        N_i^- \propto F_i+(x_-^2 + y_-^2)F_{-i}+2\sqrt{F_i F_{-i}}(c_ix_- + s_iy_-)\\
        N_i^+ \propto F_{-i}+(x_+^2 + y_+^2)F_{i}+2\sqrt{F_i F_{-i}}(c_ix_+ - s_iy_+)
    \end{split}
    \label{eq:yields}
\end{equation}

\begin{equation}
    x_{\xi}^{D^{(*)}\pi}, y_{\xi}^{D^{(*)}\pi}=\textrm{Re,Im}[\frac{r_B^{D^{(*)}\pi}e^{i\delta_B^{D^{(*)}\pi}}}{r_B^{D^{(*)}K}e^{i\delta_B^{D^{(*)}K}}}]
    \label{eq:cpodstpi}
\end{equation}
\begin{equation}
    x_{\pm}^{D^{(*)}\pi} = x_{\xi}^{D^{(*)}\pi}x^{D^{(*)}K}_{\pm} - y_{\xi}^{D^{(*)}\pi}y_{\pm}^{D^{(*)}K}, \ \ \ \ 
    y_{\pm}^{D^{(*)}\pi} = x_{\xi}^{D^{(*)}\pi}y^{D^{(*)}K}_{\pm} + y_{\xi}^{D^{(*)}\pi}x_{\pm}^{D^{(*)}K}
    \label{eq:cpodktodpi}
\end{equation}

\section{\texorpdfstring{$B^0\rightarrow DK^{*0}, D\rightarrow \KS h^+h^-$}{B0->DK*0, D->KShh}}
\begin{figure}
    \centering
        \includegraphics[width=0.33\textwidth]{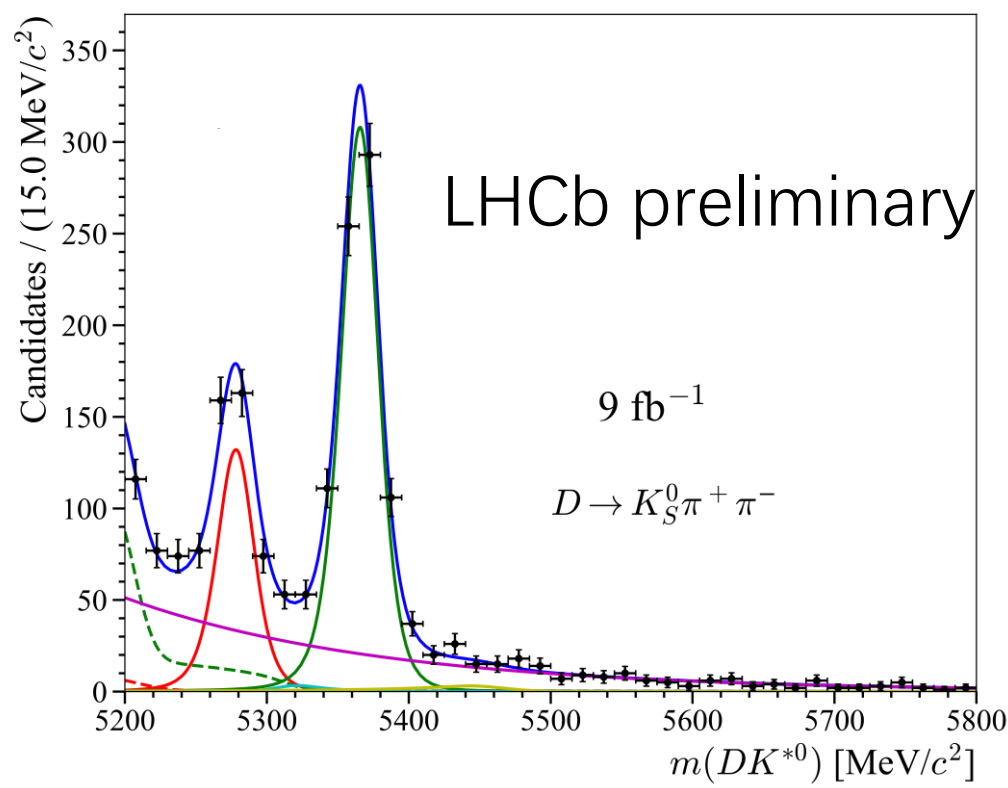}
        \includegraphics[width=0.4\textwidth]{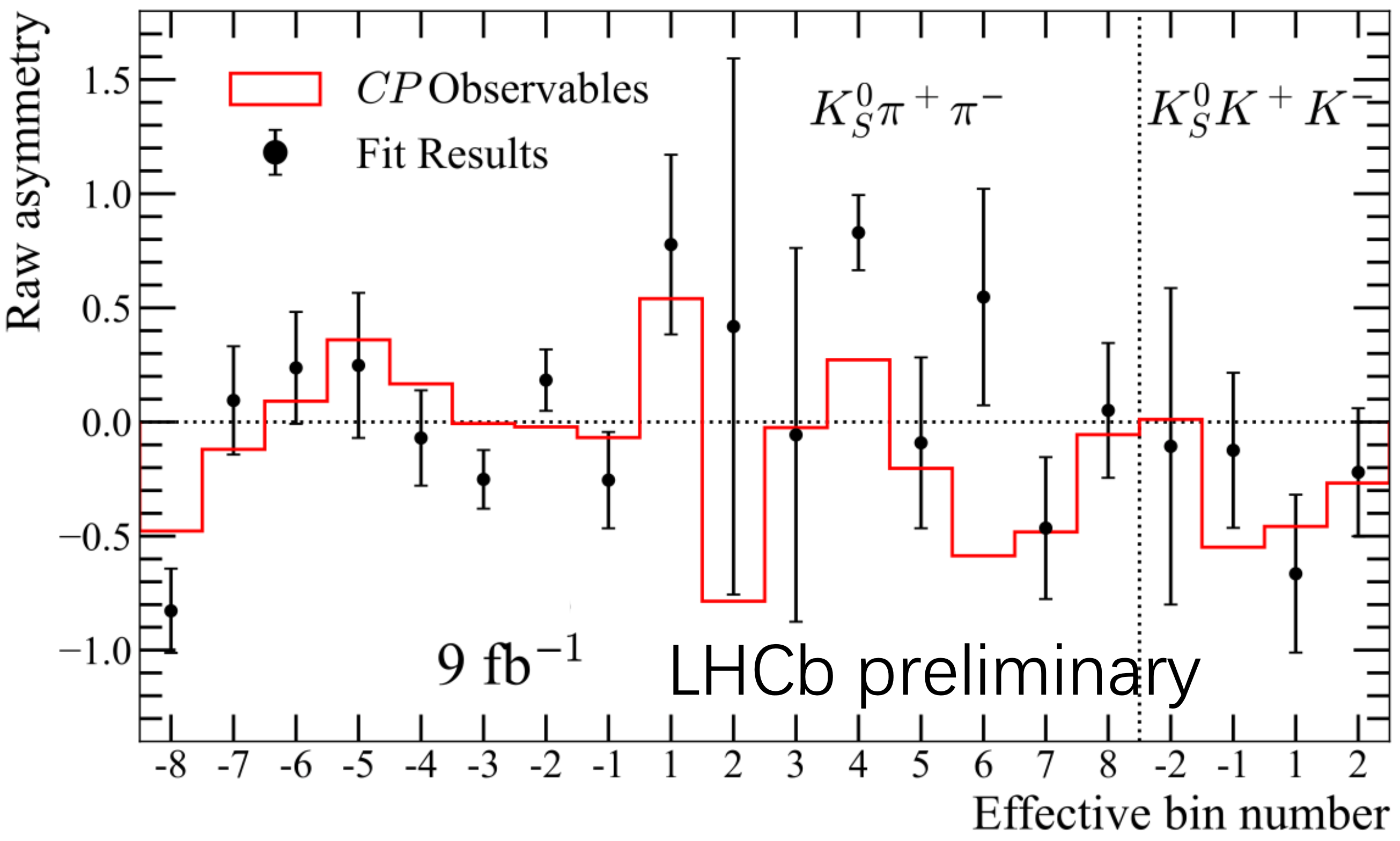}
    \caption{(left) Global fit is performed for the $B^0\rightarrow DK^{*0}$ candidates not across the DP bins of $D\rightarrow \KS \pi^+\pi^-$ decays. (right) Bin asymmetries are calculated by using the binned yields obtained from the default fit and those determined in an alternative fit where the signal yield in each bin of the DP is a free parameter.}
    \label{fig:b2dkstglobalfit}
\end{figure}

A model-independent measurement of $\CP$ violation in $B^0 \rightarrow D K^{*0}$ decays is performed. 
The DP of the self-conjugate $D \rightarrow \KS h^+ h^-$ decays is divided into bins. 
Signal yields are related to the $\CP$ observables through Eq.\ref{eq:yieldsdkst}, where the coherence factor $\kappa$ accounts for the contribution from $B^0$ decays that contain no $K^{*0}$ meson.
A global fit is performed not across the DP bins in Fig.~\ref{fig:b2dkstglobalfit}. 
\begin{equation}
    \begin{split}
        N_i^- \propto F_i+(x_-^2 + y_-^2)F_{-i}+2\kappa\sqrt{F_i F_{-i}}(c_ix_- + s_iy_-)\\
        N_i^+ \propto F_{-i}+(x_+^2 + y_+^2)F_{i}+2\kappa\sqrt{F_i F_{-i}}(c_ix_+ - s_iy_+)
    \end{split}
    \label{eq:yieldsdkst}
\end{equation}
The fit to determine the $\CP$ observables is performed simultaneously across various categories given by the $D$ decay modes, $\KS$ daughter track types, $B$-meson flavours and DP bins. The right plot in Fig.~\ref{fig:b2dkstglobalfit} displays the asymmetries calculated using the binned yields obtained from the default fit and those determined in an alternative fit where the signal yield in each DP bin is a free parameter. The good agreement demonstrates that Eq.~\ref{eq:yields} is an appropriate model for the data. The effective bin labelled $i$ is defined to compare the yield of $B^0$ decays in a bin $i$ with the yield of $\bar{B}^0$ decays in a bin $-i$. The angle $\gamma$ is determined to be $(49\pm20)^{\circ}$. This result is in good agreement with the current average from the \lhcb experiment~\cite{LHCb:2022awq} and with the previous measurement~\cite{LHCb:2016bxi}.

\section{\texorpdfstring{$B^{\pm}\rightarrow [D\pi^0/ \gamma]_{D^*}h^{\pm}, D\rightarrow \KS h^+h^-$}{Bpm->[Dpi0/gamma]_D*hpm, D->KShh}}
The $B^{\pm}\rightarrow D^*h^{\pm}$ channel is a crucial channel for the combination of the $\gamma$ measurements. Signal yields are related to $\CP$ observables through Eq.~\ref{eq:yieldsdsth}. The factor $f_{D^*}$ takes into account the $\pi$ phase difference between the $D^* \rightarrow D\pi^0$ and $D^* \rightarrow D\gamma$ decays. The 2D mass fit is performed to extract the signal from the background. 

\begin{equation}
    \begin{split}
        N_i^- \propto F_i+(x_-^2 + y_-^2)F_{-i}+2f_{D^*}\sqrt{F_i F_{-i}}(c_ix_- + s_iy_-)\\
        N_i^+ \propto F_{-i}+(x_+^2 + y_+^2)F_{i}+2f_{D^*}\sqrt{F_i F_{-i}}(c_ix_+ - s_iy_+)
    \end{split}
    \label{eq:yieldsdsth}
\end{equation}

\begin{figure}
    \centering
    \includegraphics[width=0.23\textwidth]{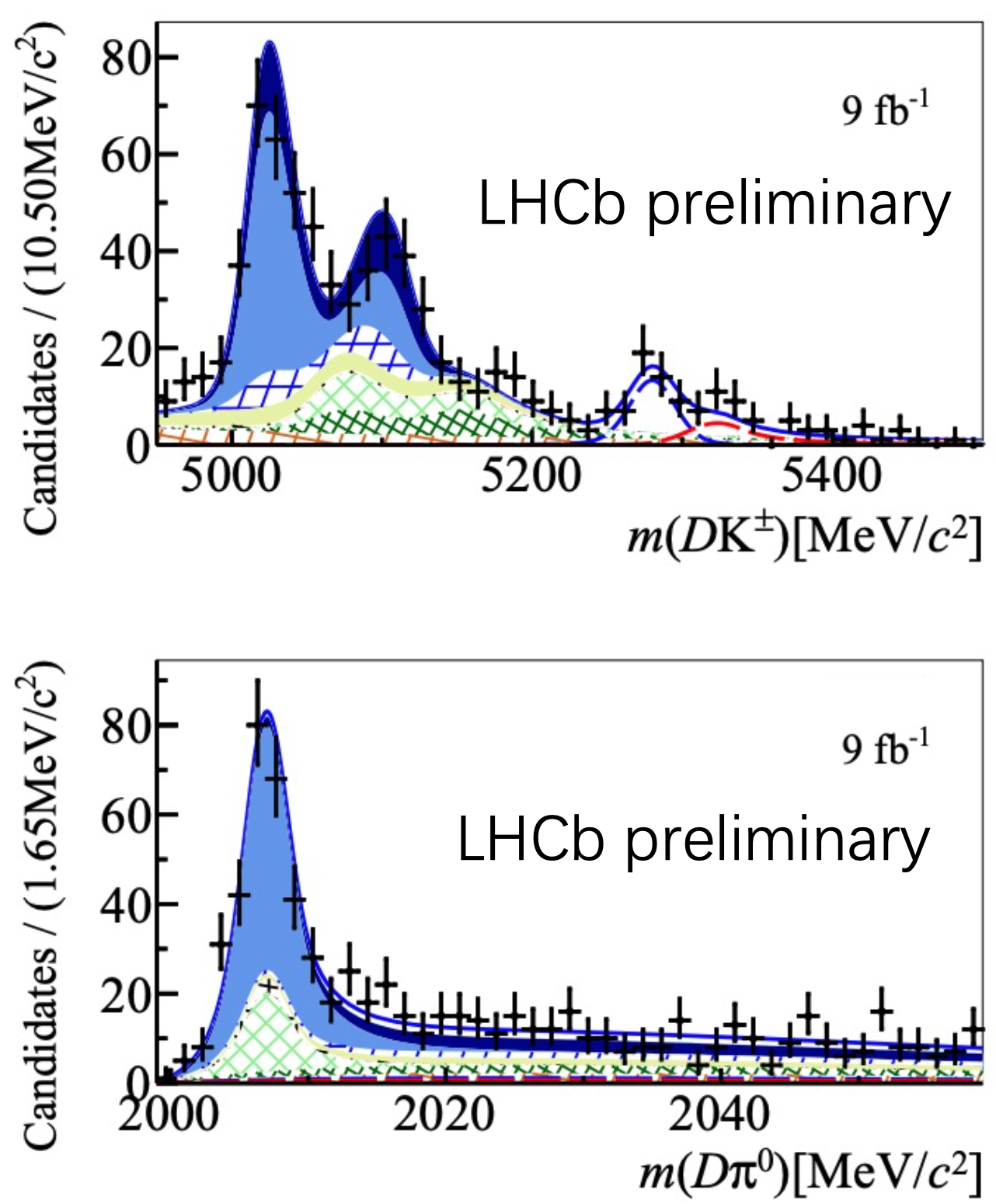}
    \includegraphics[width=0.23\textwidth]{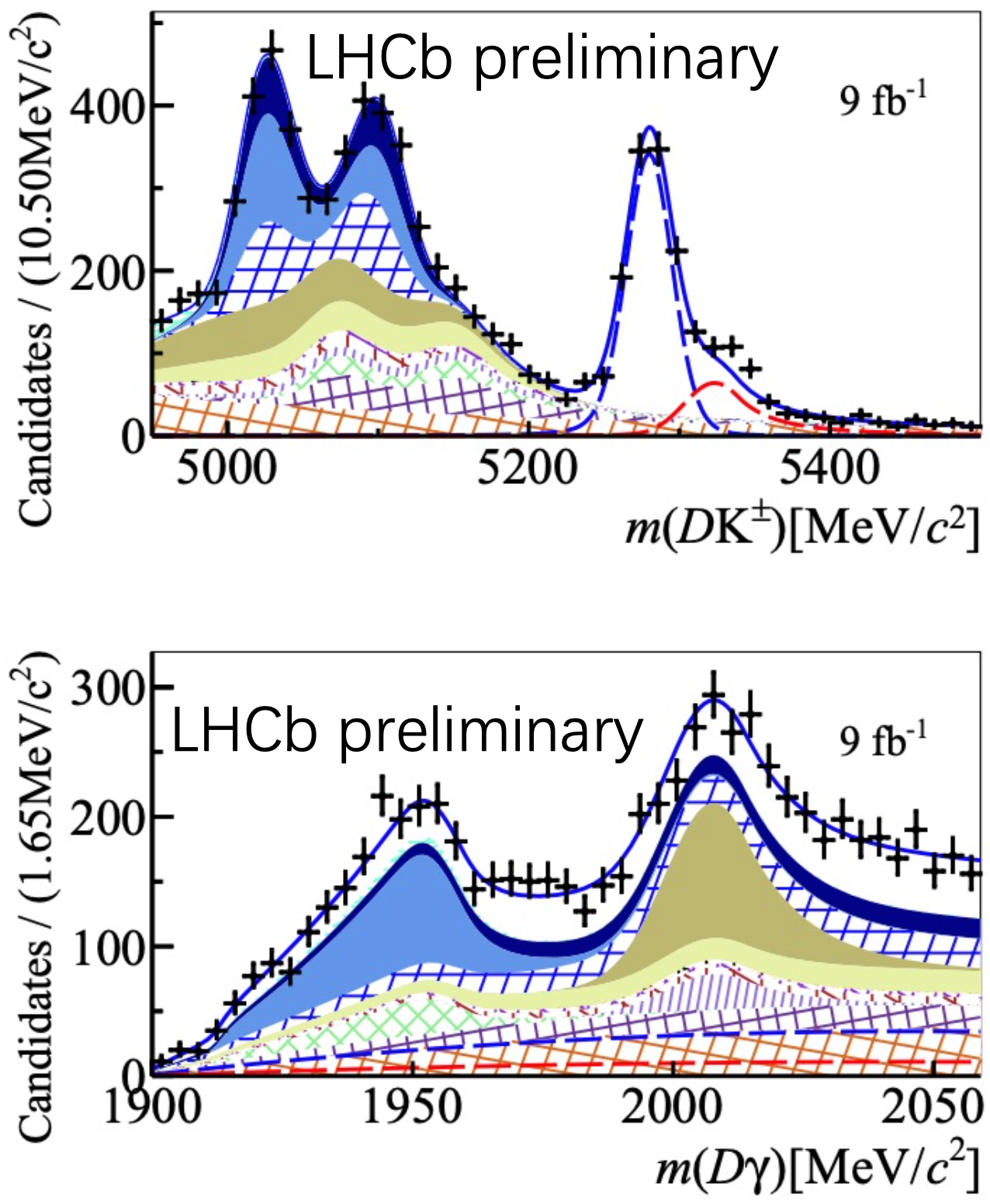}
    \includegraphics[width=0.48\textwidth]{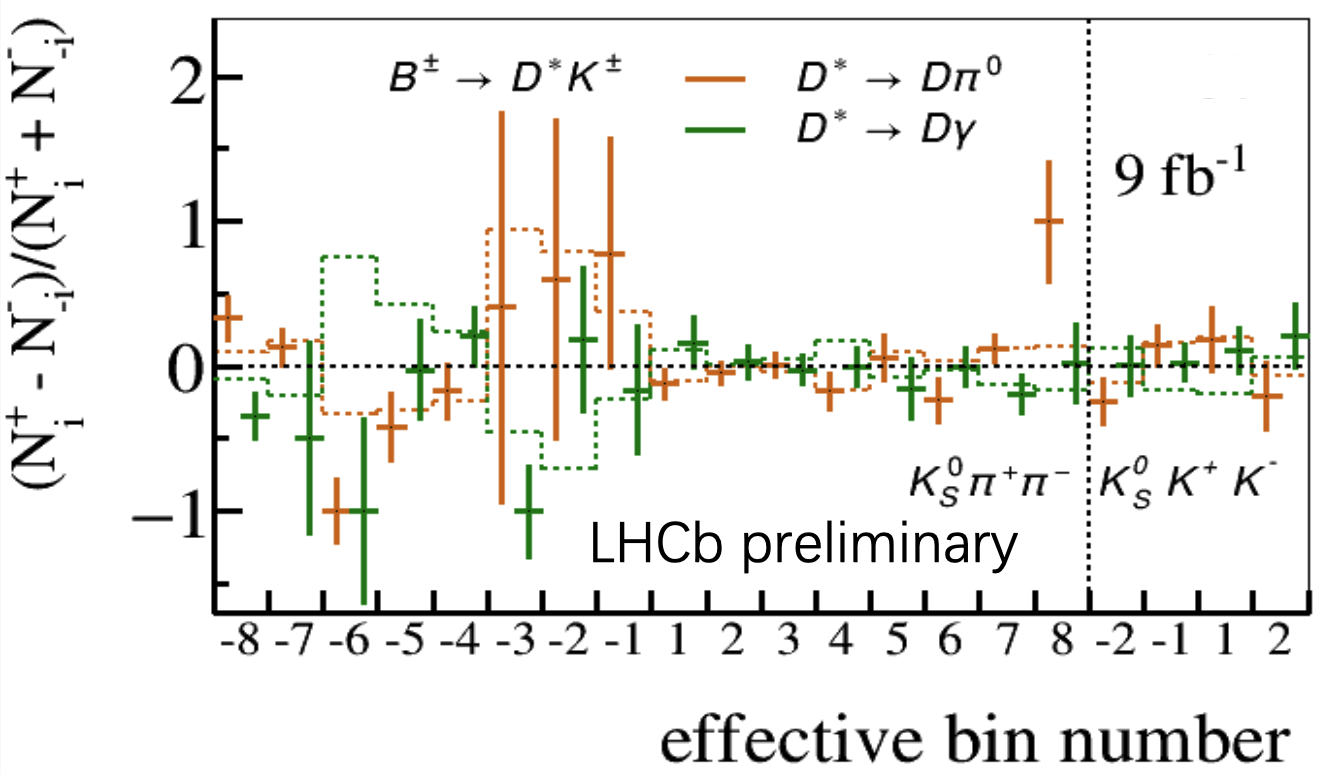}
    \caption{(left) Fit is performed simultaneously across the DP bins of the $D\rightarrow \KS \pi^+\pi^-$ decay mode and plotted with $B^{\pm}$ and DP bins merged. (right) Bin asymmetries are calculated using the binned yields obtained from the default fit and those determined in an alternative fit where the signal yield in each DP bin is a free parameter.}
    \label{fig:dsthfit}
\end{figure}

\begin{figure}
    \centering
    \includegraphics[width=0.25\textwidth]{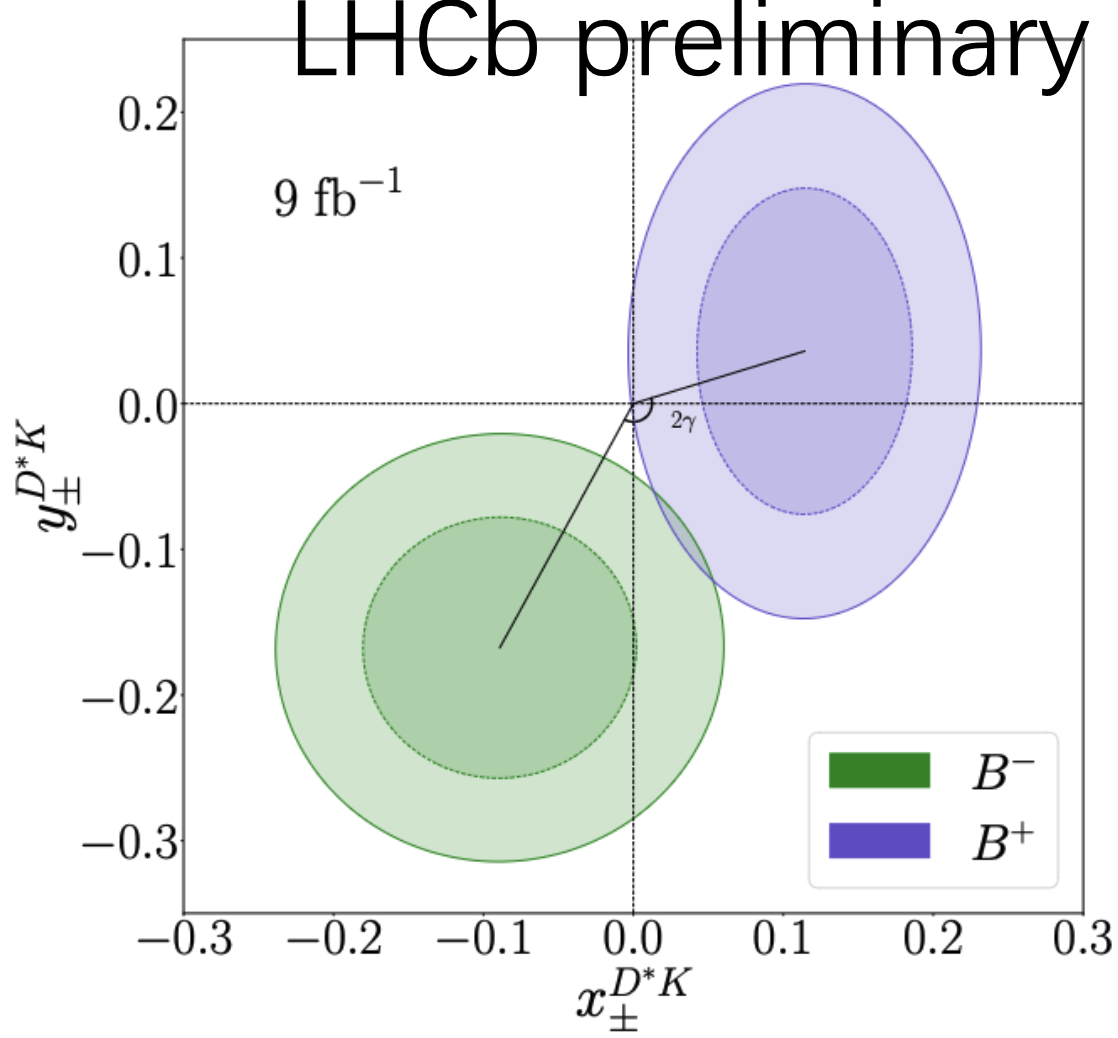}
    \includegraphics[width=0.25\textwidth]{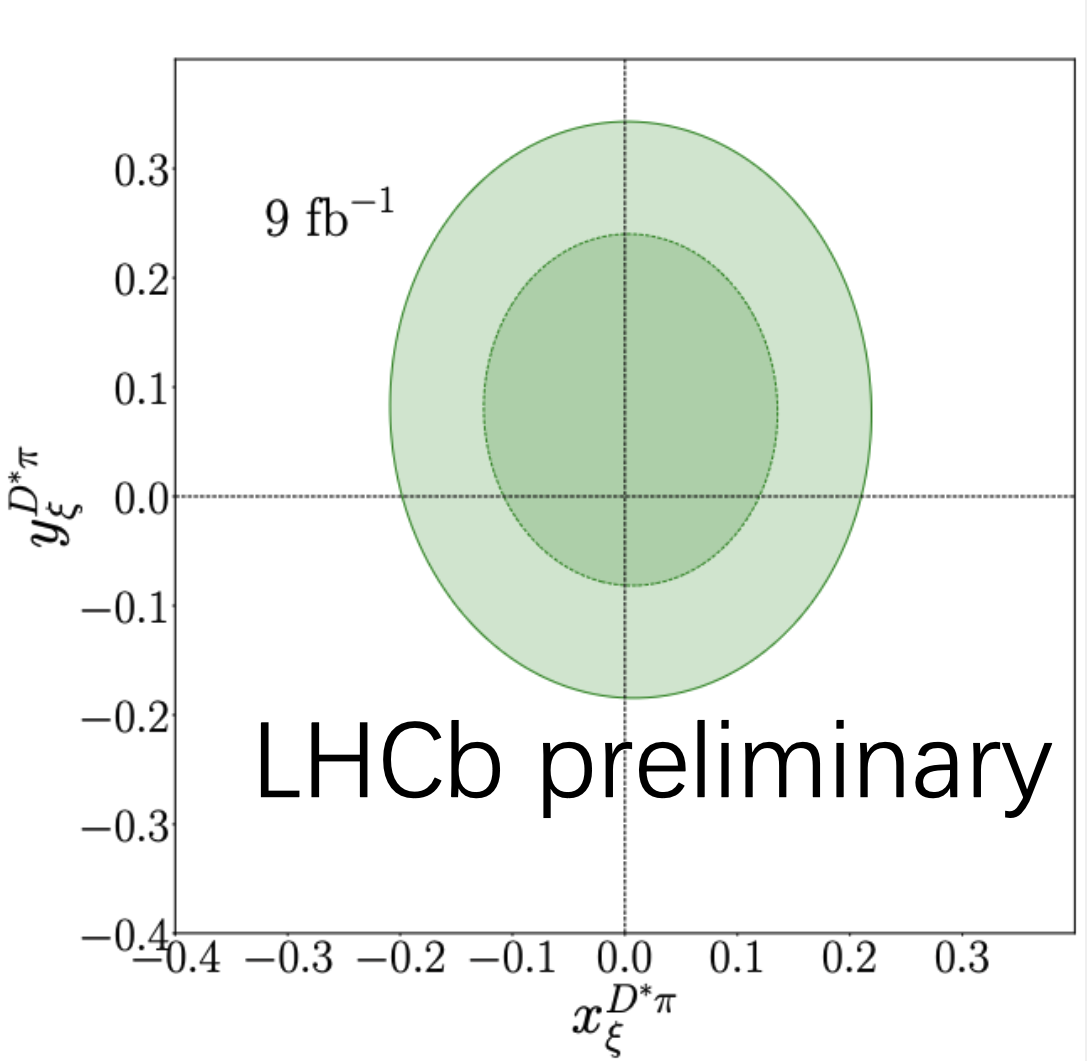}
    \includegraphics[width=0.3\textwidth]{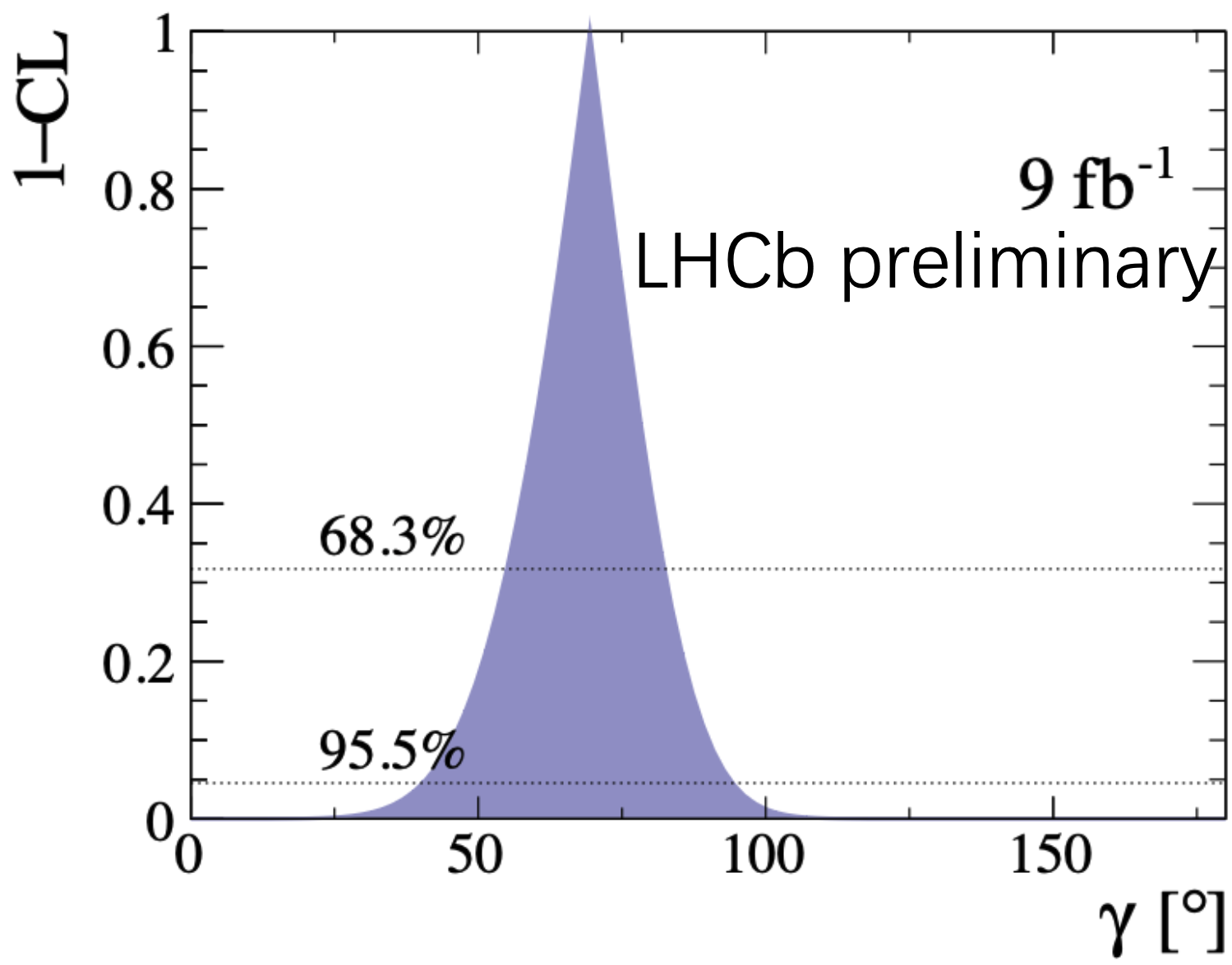}
    \caption{Confidence levels of the $\CP$ observables, (left) $x_{\pm}, y_{\pm}$, as well as (middle) $x_{\xi}, y_{\xi}$, are displayed. (right) Confidence level of $\gamma$ is shown.}
    \label{fig:b2dsthCPgamma}
\end{figure}

The fit is performed on the data simultaneously across various categories given by $B$ flavours, $D$ decays, $D^*$ decays and $D$ DP bins, which allows for the extracting of $\CP$ observables. Both fully reconstructed and partially reconstructed $D^*$ decays contribute to the $\gamma$ measurement. These components are represented by solid colours in Fig.~\ref{fig:dsthfit}. The $\CP$ observables that are measured are shown on the left and middle panels in Fig.\ref{fig:b2dsthCPgamma}. 

The right plot in Fig.~\ref{fig:dsthfit} shows the bin asymmetries calculated using the binned yields obtained from the default fit and an alternative fit where the signal yield in each DP bin is a free parameter. The good agreement indicates that Eq.\ref{eq:yieldsdsth} is an appropriate model for the data. The bin asymmetries between $D\pi^0$ and $D\gamma$ are observed to be opposite in sign. It is attributed to the $\pi$ phase difference between the $\pi^0$ mode and the $\gamma$ mode.

The $\CP$ observables are interpreted in terms of $\gamma$, which is found to be $(69^{+13}_{-14})^{\circ}$, as shown on the right in Fig.\ref{fig:b2dsthCPgamma}. This result is consistent with the world average value~\cite{Brod:2013sga} and the most precise determination using this channel. 

\section{\texorpdfstring{$B^{\pm}\rightarrow [h'^+h'^-\pi^+\pi^-]_Dh^{\pm}$}{Bpm->[h'h'pipi]_Dhpm}}

The first measurement of $\CP$ violation in the $B^{\pm} \rightarrow [K^+ K^- \pi^+ \pi^-]_D h^{\pm}$ mode, where $h$ can be either a $K$ or $\pi$, is presented~\cite{LHCb:2023yjo}. 
It includes the first determination of global $\CP$ asymmetries in this decay and updated global measurements for the $B^{\pm} \rightarrow [\pi^+\pi^- \pi^+\pi^- ]_{D} h^{\pm}$ mode. The $\CP$ observables obtained from the phase space integrated measurements can be interpreted in terms of $\gamma$ and the strong phase differences. These results are consistent with measurements using other decay channels. Moreover, the results supersede the previous measurement of \mbox{$B^{\pm} \rightarrow [\pi^+\pi^- \pi^+\pi^- ]_{D} h^{\pm}$}~\cite{LHCb:2020hdx}. 

The analysis is performed across DP bins, which are optimized for sensitivity to local $\CP$ asymmetries. External information on charm-decay parameters is required, which is currently obtained from an amplitude analysis of \lhcb data but can be updated in the future when direct measurements become available. That will allow the $\CP$ observables to be determined in a model-independent fashion. A model-dependent value of $\gamma = (116\pm12)^{\circ}$ is obtained. This result will evolve when the $\CP$ observables are determined model-independently. The precision is limited by the sample size and it is expected to improve with more future data from \lhcb.

\section{Summary}
The combination of $\gamma$ at \lhcb has been updated to include the CKM $\gamma$ measurements at \lhcb published in 2022.
There are new measurements of $\gamma$ with \mbox{$B^0\rightarrow [\KS h^+h^-]_DK^{*0}$} and \mbox{$B^{\pm}\rightarrow [D\pi^0/\gamma]_{D^*}h^{\pm}, D\rightarrow \KS h^+h^-$} and \mbox{$B^{\pm}\rightarrow [h'^+h'^-\pi^+\pi^-]_Dh^{\pm}$} decays, which will be included in the future $\gamma$ combination. 
Direct measurements of $\gamma$ in $B$ decays improve the precision of $\gamma$ angle at $\lhcb$. The uncertainty of $\gamma$ is smaller than $4^{\circ}$. Further improvement is expected with other decays and more knowledge of charm hadronic parameters.

\addcontentsline{toc}{section}{References}
\bibliographystyle{JHEP}
\bibliography{skeleton}

\end{document}